\documentclass[lettersize,journal]{IEEEtran}
\usepackage{amsmath,amsfonts}
\include{pythonlisting}
\usepackage[table,xcdraw]{xcolor}
\usepackage{algorithmic}
\usepackage{algorithm}
\usepackage[algo2e,ruled,vlined]{algorithm2e}
\usepackage{array}
\usepackage{textcomp}
\usepackage{stfloats}
\usepackage{url}
\usepackage{verbatim}
\usepackage{graphicx}
\usepackage{cite}
\usepackage{url} 
\usepackage{soul}
\usepackage{multirow}
\usepackage{nicematrix}
\usepackage{subfigure} 
\usepackage{xcolor}
\usepackage{colortbl}
\usepackage{subcaption}
\usepackage{mathrsfs}
\usepackage[colorlinks,citecolor=blue]{hyperref}

\begin{document}

\title{Augmented Intelligence in Smart Intersections: Local Digital Twins-Assisted Hybrid Autonomous Driving}

\author{Kui~Wang,
      Kazuma~Nonomura,
      Zongdian~Li,
      Tao~Yu,
      Kei~Sakaguchi,
      Omar~Hashash,
      Walid~Saad,
      Changyang~She,
      and~Yonghui~Li.}

\maketitle
\thispagestyle{plain} 

\begin{abstract}

Vehicle-road collaboration is a promising approach for enhancing the safety and efficiency of autonomous driving by extending the intelligence of onboard systems to smart roadside infrastructures. The introduction of digital twins (DTs), particularly local DTs (LDTs) at the edge, in smart mobility presents a new embodiment of augmented intelligence, which could enhance information exchange and extract human driving expertise to improve onboard intelligence. This paper presents a novel LDT-assisted hybrid autonomous driving system for improving safety and efficiency in traffic intersections. By leveraging roadside units (RSUs) equipped with sensory and computing capabilities, the proposed system continuously monitors traffic, extracts human driving knowledge, and generates intersection-specific local driving agents through an offline reinforcement learning (RL) framework. When connected and automated vehicles (CAVs) pass through RSU-equipped intersections, RSUs can provide local agents to support safe and efficient driving in local areas. Meanwhile, they provide real-time cooperative perception (CP) to broaden onboard sensory horizons. The proposed LDT-assisted hybrid system is implemented with state-of-the-art products, e.g., CAVs and RSUs, and technologies, e.g., millimeter-wave (mmWave) communications. Hardware-in-the-loop (HiL) simulations and proof-of-concept (PoC) tests validate system performance from two standpoints: \textit{(i)} The peak latency for CP and local agent downloading are 8.51~ms and 146~ms, respectively, aligning with 3GPP requirements for vehicle-to-everything (V2X) and model transfer use cases. Moreover, \textit{(ii)} local driving agents can improve safety measures by 10\% and reduce travel time by 15\% compared with conventional onboard systems. The implemented prototype also demonstrates reliable real-time performance, fulfilling the targets of the proposed system design.

\end{abstract}


\section{Introduction}

\IEEEPARstart{T}{he} emergence of autonomous driving systems demonstrates critical challenges that hinder ensuring road safety measures.This is primarily due to the necessity of processing substantial volumes of real-time data, including sensing, perception, localization, and decision-making, within stringent latency constraints \cite{liu2019edge}. Processing such data should occur within milliseconds to ensure that vehicles can react to dynamic driving environments in real time. Any delays in communications and computing could lead to incorrect or delayed decisions, which are especially dangerous in fast-evolving situations like intersections. To overcome such bottlenecks in individual autonomous vehicle systems, vehicle-road collaboration-based augmented intelligence has attracted significant interest recently (e.g., see \cite{chamola2024overtaking} and \cite{xu2024intelligent}). For instance, such augmented intelligence can facilitate information exchange between connected and automated vehicles (CAVs) and surrounding environments while also enabling the extraction of human driving expertise from dynamic and unpredictable traffic scenarios to enhance onboard intelligence. In particular, intersections in urban areas are the most hazardous as they consider the frequent presence of blind spots, vulnerable pedestrians, and stop-and-go behaviors \cite{yuan2018approach}.  A report from the US Federal Highway Administration (FHA) reveals that intersections are hotspots for vehicular accidents, accounting for 40\% of all traffic crashes and 50\% of serious collisions occurring exclusively within these areas \cite{trafficaccidents}. Hence, the development of intersection-specific augmented intelligence becomes an indispensable solution to elevate the levels of intelligence in CAVs, whereby this can contribute to enhanced levels of safety measures.

\begin{table*}[h]
\centering
\renewcommand{\arraystretch}{1.2}
\caption{Comparison with Related Works}
\begin{tabular}{cccccccc}
\hline
\hline
\multicolumn{1}{c|}{}                                 & \multicolumn{3}{c|}{\textbf{Key components}}                                       & \multicolumn{1}{c|}{}                                                                                    & \multicolumn{1}{c|}{}                                                                      & \multicolumn{1}{c|}{}                                                                                &                                                               \\ \cline{2-4}
\multicolumn{1}{c|}{\multirow{-2}{*}{\textbf{Ref.}}}           & \textbf{Cloud}   & \textbf{Vehicles} & \multicolumn{1}{c|}{\textbf{RSUs}}                            & \multicolumn{1}{c|}{\multirow{-2}{*}{\begin{tabular}[c]{@{}c@{}}\textbf{Unique traffic}\\ \textbf{features}\end{tabular}}} & \multicolumn{1}{c|}{\multirow{-2}{*}{\begin{tabular}[c]{@{}c@{}}\textbf{SAE}\\ \textbf{Level}\end{tabular}}} & \multicolumn{1}{c|}{\multirow{-2}{*}{\begin{tabular}[c]{@{}c@{}}\textbf{Validation} \\ \textbf{Methods}\end{tabular}}} & \multirow{-2}{*}{\textbf{Research objective}}                          \\ \hline
\rowcolor[HTML]{FFFC9E} 
\multicolumn{8}{c}{\cellcolor[HTML]{FFFC9E}\textbf{Cooperative perception (CP)-related studies}}                                                                                                                                                                                                                                                                                                                                                                                                                          \\ \hline
\rowcolor[HTML]{EFEFEF} 
\multicolumn{1}{c|}{\cellcolor[HTML]{EFEFEF}{[}17{]}} &         & $\surd$  & \multicolumn{1}{c|}{\cellcolor[HTML]{EFEFEF}$\surd$} & \multicolumn{1}{c|}{\cellcolor[HTML]{EFEFEF}}                                                            & \multicolumn{1}{c|}{\cellcolor[HTML]{EFEFEF}-}                                             & \multicolumn{1}{c|}{\cellcolor[HTML]{EFEFEF}Simulation}                                              & Generation of future BEV occupancy map                        \\
\multicolumn{1}{c|}{{[}18{]}}                         &         & $\surd$  & \multicolumn{1}{c|}{$\surd$}                         & \multicolumn{1}{c|}{$\surd$}                                                                             & \multicolumn{1}{c|}{-}                                                                     & \multicolumn{1}{c|}{Dataset}                                                                         & Early prediction of intersection movement and warning         \\
\rowcolor[HTML]{EFEFEF} 
\multicolumn{1}{c|}{\cellcolor[HTML]{EFEFEF}{[}19{]}} &         & $\surd$  & \multicolumn{1}{c|}{\cellcolor[HTML]{EFEFEF}$\surd$} & \multicolumn{1}{c|}{\cellcolor[HTML]{EFEFEF}$\surd$}                                                     & \multicolumn{1}{c|}{\cellcolor[HTML]{EFEFEF}-}                                             & \multicolumn{1}{c|}{\cellcolor[HTML]{EFEFEF}Dataset}                                                 & Adaptive construction of a dynamic perception representation  \\
\multicolumn{1}{c|}{{[}27{]}}                         &         & $\surd$  & \multicolumn{1}{c|}{$\surd$}                         & \multicolumn{1}{c|}{}                                                                                    & \multicolumn{1}{c|}{-}                                                                     & \multicolumn{1}{c|}{Simulation}                                                                      & Intersection traffic safety framework for collision avoidance \\ \hline
\rowcolor[HTML]{FFFC9E} 
\multicolumn{8}{c}{\cellcolor[HTML]{FFFC9E}\textbf{Cooperative driving (CD)-related studies}}                                                                                                                                                                                                                                                                                                                                                                                                                             \\ \hline
\rowcolor[HTML]{EFEFEF} 
\multicolumn{1}{c|}{\cellcolor[HTML]{EFEFEF}{[}14{]}} &         & $\surd$  & \multicolumn{1}{c|}{\cellcolor[HTML]{EFEFEF}$\surd$} & \multicolumn{1}{c|}{\cellcolor[HTML]{EFEFEF}}                                                            & \multicolumn{1}{c|}{\cellcolor[HTML]{EFEFEF}-}                                             & \multicolumn{1}{c|}{\cellcolor[HTML]{EFEFEF}Simulation}                                              & Federated learning framework to reduce idling time            \\
\multicolumn{1}{c|}{{[}15{]}}                         &         &          & \multicolumn{1}{c|}{$\surd$}                         & \multicolumn{1}{c|}{}                                                                                    & \multicolumn{1}{c|}{-}                                                                     & \multicolumn{1}{c|}{Simulation}                                                                      & Traffic signal control system to reduce idling time           \\
\rowcolor[HTML]{EFEFEF} 
\multicolumn{1}{c|}{\cellcolor[HTML]{EFEFEF}{[}29{]}} &         & $\surd$  & \multicolumn{1}{c|}{\cellcolor[HTML]{EFEFEF}$\surd$} & \multicolumn{1}{c|}{\cellcolor[HTML]{EFEFEF}}                                                            & \multicolumn{1}{c|}{\cellcolor[HTML]{EFEFEF}L4/L5}                                         & \multicolumn{1}{c|}{\cellcolor[HTML]{EFEFEF}Indoor robots}                                           & Intersection coordination framework for trajectory planning   \\ \hline
\rowcolor[HTML]{FFFC9E} 
\multicolumn{8}{c}{\cellcolor[HTML]{FFFC9E}\textbf{Digital Twin (DT)-based studies}}                                                                                                                                                                                                                                                                                                                                                                                                                                      \\ \hline
\rowcolor[HTML]{EFEFEF} 
\multicolumn{1}{c|}{\cellcolor[HTML]{EFEFEF}{[}40{]}} &         & $\surd$  & \multicolumn{1}{c|}{\cellcolor[HTML]{EFEFEF}$\surd$} & \multicolumn{1}{c|}{\cellcolor[HTML]{EFEFEF}}                                                            & \multicolumn{1}{c|}{\cellcolor[HTML]{EFEFEF}-}                                             & \multicolumn{1}{c|}{\cellcolor[HTML]{EFEFEF}Field test}                                              & Highway DT for CAV perception enhancement                     \\
\multicolumn{1}{c|}{{[}42{]}}                         & $\surd$ & $\surd$  & \multicolumn{1}{c|}{}                                & \multicolumn{1}{c|}{}                                                                                    & \multicolumn{1}{c|}{L2}                                                                    & \multicolumn{1}{c|}{Field test}                                                                      & DT-based ramp merging framework                               \\
\rowcolor[HTML]{EFEFEF} 
\multicolumn{1}{c|}{\cellcolor[HTML]{EFEFEF}{[}43{]}} & $\surd$ & $\surd$  & \multicolumn{1}{c|}{\cellcolor[HTML]{EFEFEF}}        & \multicolumn{1}{c|}{\cellcolor[HTML]{EFEFEF}}                                                            & \multicolumn{1}{c|}{\cellcolor[HTML]{EFEFEF}L2}                                            & \multicolumn{1}{c|}{\cellcolor[HTML]{EFEFEF}Simulation}                                              & DT-assisted FIFO framework at intersections                   \\ \hline
\multicolumn{1}{c|}{This work}                        & $\surd$ & $\surd$  & \multicolumn{1}{c|}{$\surd$}                         & \multicolumn{1}{c|}{$\surd$}                                                                             & \multicolumn{1}{c|}{L4/L5}                                                                 & \multicolumn{1}{c|}{Field test}                                                           & Hybrid autonomous driving system with local strategies        \\ \hline
\hline
\end{tabular}
\end{table*}

With recent advances in digital twin (DT) technologies, the concept of mobility digital twins (MDTs) serves as an exemplary \textit{embodiment of augmented intelligence}, offering a robust platform to address various on-road challenges \cite{yang2023parallel, hu2024how, yu2023internet}. MDTs provide a multi-layer framework that integrates real-time data acquisition, simulation, and predictive modeling to mirror the behaviors of physical systems within a virtual environment. In the context of autonomous driving, MDTs can capture the intricacies of traffic dynamics, road infrastructure, and vehicle behaviors to create a comprehensive model that continuously adapts to real-world conditions\cite{irfan2024towards}. On the one hand, MDTs encompass comprehensive global traffic information, thereby providing connected users with far-reaching, computation-intensive services such as global vehicle routing that extend well beyond immediate sensory capabilities \cite{teng2023motion, wang2024smart}, which enables autonomous driving systems to make smarter decisions based on system-level insights derived from real-time global traffic data. On the other hand, MDT incorporates distributed intelligence across various smart entities \cite{tihanyi2021towards}, including CAVs, smart traffic signals, and roadside units (RSUs). This distributed framework facilitates the provision of localized, delay-sensitive services. By leveraging real-time data and computational power within these interconnected smart entities, MDTs can dynamically adapt to the specific conditions of different traffic areas, ensuring more responsive and effective decision-making for CAVs.

Therefore, we introduce a novel concept named ``local DT (LDT)", a specialized DT system designed to monitor, analyze, and optimize traffic behaviors within specific areas, i.e., traffic intersections. Acting as a subsystem of a global DT (GDT), the LDT enhances flexibility and reusability by leveraging local services, storage, models, and edge connections to ensure faster response \cite{kondo2024industrial}. In contrast to GDT, which focuses on large-scale traffic patterns, LDTs are suitable to deliver low-latency responses by processing data locally at the edge, enabling real-time decision-making for time-sensitive tasks such as vehicle maneuvering, intersection management, and adaptive traffic signal timing (e.g., see \cite{chou2024novel} and \cite{chu2022traffic}). In addition, this localized approach allows the system to address the intricate and dynamic nature of intersections more effectively than a generalized global system. Considering the unique characteristics of each traffic intersection, including its road topology, the volume of traffic, and the diversity of traffic participants, LDTs can generate strategies tailored to each intersection scenario. The overarching goal of this paper is to propose an LDT-assisted hybrid vehicle maneuvering system in intersection areas that explores the potential revolution of autonomous driving systems brought by vehicle-road collaboration-based augmented intelligence. Through this system, we aim to enhance the decision-making capabilities of CAVs with LDT-based cooperative perception (CP) and local driving strategies provided by RSUs, enabling them to navigate vehicles through intersections more safely and efficiently.

\subsection{Related Works}

Existing vehicle-road collaboration technologies can be broadly categorized into two types: CP and cooperative driving (CD). CP allows a network to leverage advancements in intelligent transportation systems (ITS) and vehicle-to-everything (V2X) technologies to improve the situational awareness capability of CAVs by reducing the occurrence of safety-critical events. For instance, CAVs can obtain more road traffic information via vehicle-to-infrastructure (V2I) communications which allows them to make decisions more reasonably. Hence, various studies have focused on enhancing CAV operations by using CP \cite{sakaguchi2021towards, chang2023bev, komol2023deep, tan2024dynamic, huang2024toward, shan2020demonstrations, cai2023consensus, meng2023hydro, xiao2023overcoming, ren2024interruption, wu20233d, Shahriar2023enhancing}. For example, the work in \cite{sakaguchi2021towards} proposes raw sensor data sharing to support automated driving via millimeter-wave (mmWave) V2X. The authors in \cite{chang2023bev} propose a birds-eye-view (BEV) grid occupancy prediction method to support various applications like driving safety warning and cooperative traffic control. Moreover, the work in \cite{komol2023deep} and \cite{tan2024dynamic} also notices the importance of unique features of different intersections, where intersection-specific strategies are deployed for driver-intended movements prediction and road-to-vehicle visual information sharing, respectively. Several works such as \cite{huang2024toward} and \cite{shan2020demonstrations} also highlight the practical aspects of CP system implementation and deployment. In our previous work \cite{li2023het}, a software-defined vehicular network (SDVN) architecture for heterogeneous V2X is implemented to realize CP and ensure safe autonomous driving.

On the other hand, CD allows CAVs to coordinate their maneuvers in complex traffic environments, such as intersections and highways, thereby improving overall traffic safety and efficiency. Research in this area mainly focuses on developing sophisticated algorithms and robust communications protocols that enhance the collaborative decision-making capabilities of CAVs in dynamic and uncertain environments \cite{zhang2021trajectory, yang2021cooperative, xu2023multi, wang2019cooperative, wang2023stop, yu2022cd, zhang2023bi, liu2023safety, Charouh2022video, wang2023data}. For instance, multi-vehicle coordination systems utilize real-time data sharing and predictive modeling to enable vehicles to anticipate the actions of others and adjust their trajectories accordingly, thereby reducing the risk of collisions and improving traffic flow \cite{zhang2021trajectory, yang2021cooperative, xu2023multi}. Advanced intersection management models integrate traffic signal optimization and vehicle trajectory planning, considering factors such as speed, density, and pedestrian movements, to minimize congestion and enhance safety \cite{wang2019cooperative, wang2023stop}. 

In recent years, the introduction of DT in the transportation systems has become a hot topic in CD and CP fields as the synergistic integration between the physical and cyber worlds allows for enhanced real-time decision-making and predictive analytics \cite{irfan2024towards, zheng2023opencda, liao2021cooperative, buechel2019fortuna, krammer2019providentia, wang2021digital}. For example, in \cite{buechel2019fortuna} and \cite{krammer2019providentia}, authors propose and implement transportation DT systems to improve automated driving functions like environmental perception. The work in \cite{wang2021digital} focuses on the DT-based first-in-first-out (FIFO) slot reservation algorithm in traffic intersections to reduce travel time and energy consumption. In our previous work \cite{wang2024smart}, an MDT system in the real world is established to support cooperative global route planning for CAVs. Collectively, these advancements highlight the potential of vehicle-road collaboration to significantly augment the onboard intelligence of autonomous vehicles, paving the way for safer and more efficient traffic systems.

However, most existing research in CP and CD fields primarily focuses on the sharing of dynamic information, such as vehicle states, predicted behaviors, and decisions, as shown in Table I. Indeed, dynamic information is crucial for operational safety, but introducing \textit{(transient) static information and inherent unique road traffic features}, and based on which, generating \textit{local strategies} have great potential to transformatively boost the development of autonomous driving. For example, a human driver commuting daily past an intersection near a kindergarten will quickly learn to pay extra attention and adopt a more careful and conservative \textit{local strategy}. How to enable CAVs to gain such knowledge and \textit{local strategies} is the core motivation of this research. Additionally, most related studies employ validation methods of simulation or collected datasets, lacking implementation and outdoor tests to validate the system effectiveness in the real world. Hence, validating the benefits of \textit{local strategy} with real-world experiments is also a key consideration in this research.

\subsection{Contributions}
RSUs offer an excellent solution to this challenge. With their fixed positions and continuous sensing capabilities, RSUs can monitor the driving behaviors within certain coverage areas. In this study, we also endow RSU with edge computing and V2X communications, so RSU serves as the infrastructure of LDT that can share perception information with surrounding vehicles, and learn and accumulate local human driving knowledge over time. In this setting, there is no real-time interaction between vehicle agents and the traffic environment. To achieve this goal, we propose an innovative RSU-based offline reinforcement learning (RL) framework to enable offline model training and online inference for hybrid vehicle control. Hardware-in-the-loop (HiL) simulations and proof-of-concept (PoC) field trial results demonstrate that the proposed system offers a substantial enhancement in traffic safety and efficiency within RSU-installed areas, outperforming conventional autonomous driving systems. The main contributions of this paper are summarized as follows:
\begin{itemize}
    \item \emph{System design}: We propose a high-level system design for hybrid autonomous driving with cloud/edge computing. Based on the standards from the 3rd generation partnership project (3GPP), we analyze system requirements considering communications issues.
    
    \item \emph{Offline RL framework}: Based on V2I communications, an offline RL framework, including an environment perception data encoder and a kinematic-model-based decoder, is designed across RSU and CAV edge systems.
    
    \item \emph{Implementation and field trial}: We implement two RSUs with LiDAR sensors, edge computing, and V2X communications modules. Then, hybrid autonomous driving HiL and PoC experiments are conducted to validate the proposed system.
    
    \item \emph{Safety and efficiency improvements}: Compared to conventional autonomous driving systems, our proposed system demonstrates significant improvements with safety enhancement of about 10\% and travel time reduction of about 15\%. The implemented system also aligns with 3GPP requirements and demonstrates reliable real-time performance, fulfilling the targets of the proposed design.

\end{itemize}

The rest of this paper is organized as follows. Section II explains the system architecture of LDT-assisted hybrid autonomous driving system, and then we analyze the requirements for proposed system considering communications issues. The detailed approach, i.e., the offline RL framework, as well as algorithm design, are presented in Section III. Section IV presents the HiL and PoC experiment setups. In Section V, test and evaluation results, including model training, empirical metrics measurement, and driving performance are shown and discussed. Finally, we draw conclusions in Section VI.

\section{LDT-Assisted Hybrid Autonomous Driving}
\begin{figure*}[!t]
    \centerline{\includegraphics[width=0.82\textwidth]{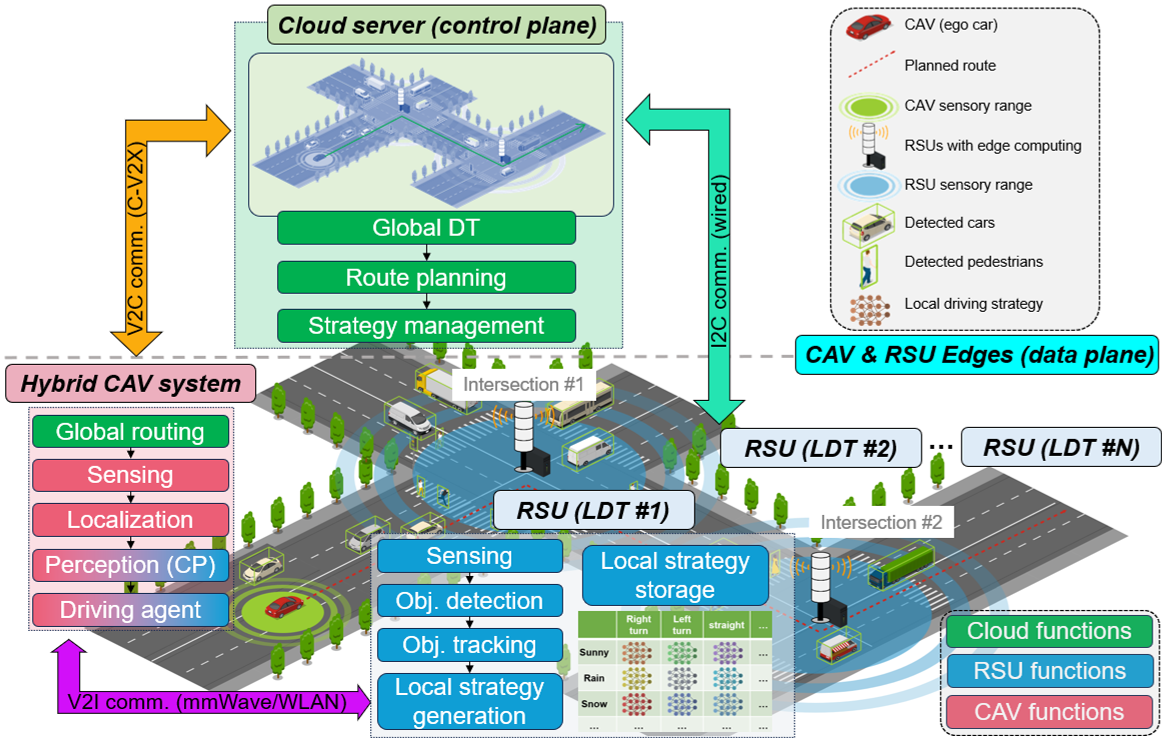}}
    \caption{LDT-based hybrid autonomous driving system architecture.}
    \label{fig: sys}
\end{figure*}

In this section, we introduce the architecture of our LDT-assisted hybrid autonomous driving system, leveraging computational resources at the central cloud and edge (RSUs and CAVs), as shown in Fig.~\ref{fig: sys}. The technical requirements of the proposed system are discussed based on the 3GPP standards for SSMS and AI/ML-based automotive networked systems. 

\subsection{Overall System Design}
The proposed LDT-assisted hybrid autonomous driving system aims to enhance road safety and the driving efficiency of autonomous vehicles at intersections, which are fraught with risks due to frequent pedestrian crossings and blind spots, posing challenges for autonomous vehicles. In our system, we assume that RSUs are deployed in complex traffic environments, including non-signalized intersections. They are equipped with various sensors and computing units. In addition, we assume that the CAVs (i.e., target users) in the proposed system have achieved high-level automation with all necessary functional blocks, including sensing, localization, environmental perception, motion planning, and motion control \cite{yurtsever2020survey}. A computationally capable central cloud is also deployed at the top of this system. The detailed functionalities of RSUs, CAVs, and the central cloud are discussed next:

\emph{1) RSU edge:} 
The RSUs persistently monitor the traffic environment using mounted sensors and advanced objection detection techniques. In this system, RSUs are the hosts of LDTs. They perform three critical tasks: \textit{(i)} upload real-time processed perception information to the cloud; \textit{(ii)} share real-time perception information with surrounding CAVs; \textit{(iii)} record, analyze, and extract knowledge from the driving behaviors of passing-by vehicles, then generate \emph{local driving strategies} tailored for locational intersections within specific environmental conditions, such as varying times of day and weather patterns.

\emph{2) Central cloud:} The central cloud establishes a real-time GDT, after receiving and synchronizing perception data from distributed RSUs, based on which the central cloud also plans global routes for each CAV user in the traffic network \cite{wang2024smart}. Since the cloud is aware of all the intersections that the CAVs will pass through, it can determine and ``inform" them of the most appropriate \emph{local driving strategies} to guide CAVs effectively by combining real-time assessments of current conditions.

\emph{3) CAV edge:} The CAVs follow the global routes planned by the central cloud. When CAVs approach the intersections managed by RSUs and enter their V2I coverage, the CAVs start downloading \emph{local driving strategies} from RSUs and follow. Instead, they depend on their inherent autonomous driving systems while driving on road segments without RSUs. The operations of CAV systems at RSU-assisted intersections can be separated into three steps: \textit{(i)} request \emph{local driving strategies} from RSUs; \textit{(ii)} receive real-time perception data from RSUs and fuse with onboard sensing; \textit{(iii)} switch the driving agent from inherent motion planning and motion control modules to \emph{local driving strategies}. An intersection-specific hybrid autonomous driving system is thus achieved.

For example, in Fig.~\ref{fig: sys}, a CAV scheduled to traverse the intersections \#1 and \#2 will request the \emph{local driving strategies} that match current time and weather from RSU \#1 and \#2, respectively. Because such \emph{local driving strategies} are extracted and learned from local driving behaviors, adopting them allows CAVs to make more reasonable decisions tailored for corresponding local areas. Upon the CAV entering the V2I communications coverage of RSU \#1 and \#2, the CP service is provided to extend the horizon of onboard sensing. Such enhanced perception endows the CAV with a comprehensive understanding of the traffic environment and the capability to identify potential hazards concealed within blind spots.

This system design capitalizes on heterogeneous communications and computing resources on edge and cloud \cite{wang2023vtc}. The edge computing on RSUs processes raw sensor data into object-level detection information, in order to reduce the large transmission overhead and delays to the CAVs. Since vehicle control is an extremely delay-sensitive and safety-critical task, the RSUs will directly transmit their \emph{local driving strategies} through V2I, to be executed by edge computing on CAVs. In the system, \emph{local driving strategies} dissemination is managed by the control plane, which is located in the central cloud, while the data plane is at the edge.
The main reasons are that: \textit{(i)} the cloud system is responsible for planning the global route for CAVs. As a result, the cloud can prefetch the required \emph{local driving strategies} along the planned route and delegate RSU edges to transmit them in advance, and \textit{(ii)} considering the large data size of AI/ML-based models, V2I transmission, compared to V2C transmission, is much more cost-efficient and will not lead to the model reception and execution failure on CAVs due to unstable V2C delay.

\subsection{Requirements Analysis Considering Communications Issues}
The proposed system aims to enhance the safety and efficiency of CAVs by utilizing V2I communications to facilitate the download of \textit{local driving strategies} and the sharing of perception data. To achieve such functions, the system must meet certain communications requirements to ensure reliable and timely data exchange. Therefore, it is necessary to define specific key performance indicators (KPIs) that the system should fulfill. \emph{Local driving strategies} downloading and perception data sharing correspond to an AI/ML model transfer use case and a V2X use case proposed by 3GPP \cite{garcia2021tutorial, taleb2023ai}, called AI/ML based automotive networked systems and SSMS, respectively. By definition, AI/ML-based automotive networked systems frequently exchange and share AI/ML model layers in a distributed and/or federated network as determined by the system in response to a change of events, conditions, or emergency situations to improve system accuracy. SSMS enables the sharing of raw or processed sensor data to build collective situational awareness, thereby enabling safety-critical applications like intelligent intersections to ensure the safety of all road users, including pedestrians and emergency vehicles.

To assess whether the proposed hybrid autonomous driving system can meet the necessary requirements of the above two use cases, some KPIs that need to be validated are summarized in Table~\ref{tab:req}. For the AI/ML-based automotive networked system use case, some required communications KPIs are determined based on 3GPP technical report (TR) 22.874 \cite{3gpptr22874}, including data size, end-to-end (E2E) latency, and peak data rate for data downloading. In the proposed system, CAVs shall download a series of AI models of \emph{local driving strategies} for different intersections from RSUs. Here, we define data downloading as that CAV downloads only one of the AI models. The proposed system should be able to support downloading of data with a maximum size of 2$\sim$40~MB with a latency of up to 500$\sim$1000~ms and a peak data rate of up to 100~Mbps.

\begin{table}[]
\centering
\renewcommand{\arraystretch}{1.2}
\caption{Communications KPIs of AI/ML based automotive networked system and SSMS use cases}
\begin{tabular}{c|c|c}
\hline
\hline
\textbf{Use cases}                                               & \textbf{\begin{tabular}[c]{@{}c@{}}AI/ML based automotive\\ networked system\end{tabular}} & \textbf{\begin{tabular}[c]{@{}c@{}}Sensor and state map \\ sharing (SSMS)\end{tabular}} \\ \hline
\rowcolor[HTML]{EFEFEF} 
\begin{tabular}[c]{@{}c@{}}Reliability \\ (\%)\end{tabular}      & -                                                                                      & 90                                                                                          \\ \hline
\rowcolor[HTML]{FFFC9E} 
\begin{tabular}[c]{@{}c@{}}E2E latency \\ (ms)\end{tabular}      & 500$\sim$1000                                                                                      & 10                                                                                   \\ \hline
\rowcolor[HTML]{EFEFEF} 
\begin{tabular}[c]{@{}c@{}}Peak data rate \\ (Mbps)\end{tabular} & 100                                                                                      & 25                                                                                        \\ \hline
\rowcolor[HTML]{FFFC9E} 
\begin{tabular}[c]{@{}c@{}}Data size \\ (MB)\end{tabular}        & 2$\sim$40                                                                                       & -                                                                                       \\ \hline
           \hline
\end{tabular}
\label{tab:req}
\end{table}

According to 3GPP TR 22.886 \cite{3gpptr22886}, the requirements of SSMS include reliability, E2E latency, and peak data rate. The reliability refers to the probability of successful transmission, which should be higher than 90\%. The E2E latency refers to the one-way delay for delivering a piece of information from a source to a destination, without the application-layer processing delay. In this system, the E2E latency of SSMS is defined as the one-way communications delay of CP, where RSU shares real-time perception information with surrounding CAVs, and the latency should be less than 10~ms. The peak data rate should be higher than 25~Mbps.

\section{Approach: Offline RL Framework for Localized Learning}
In this section, we propose an offline RL framework to extract and learn human driving knowledge. Offline RL is commonly applied to autonomous driving systems because it breaks the constraint that agents should interact with the environment, thereby avoiding risky exploration in real driving processes \cite{levine2020offline, li2023boosting}. Unlike online RL, which requires continuous interaction between agents and the environment, offline RL allows us to safely utilize vast amounts of pre-collected driving data from human-driven vehicles to train driving models without exposing the vehicles to hazardous real-world conditions. Moreover, compared to non-RL approaches, which may struggle to generalize across diverse driving conditions and require extensive hand-crafted rules, offline RL offers a more flexible and data-driven way to learn complex driving behaviors. In our system, the role of gaining human knowledge is placed on the RSU edge, and autonomous systems in CAVs only deploy the trained agent to achieve vehicle maneuvering. In particular, we introduce a sensory data encoder, a formulation of a partially observed Markov decision process (POMDP), and a kinematic-model-based vehicle maneuvering decoder to enhance the decision-making process.

\begin{figure*}[!t]
    \centerline{\includegraphics[width=1\textwidth]{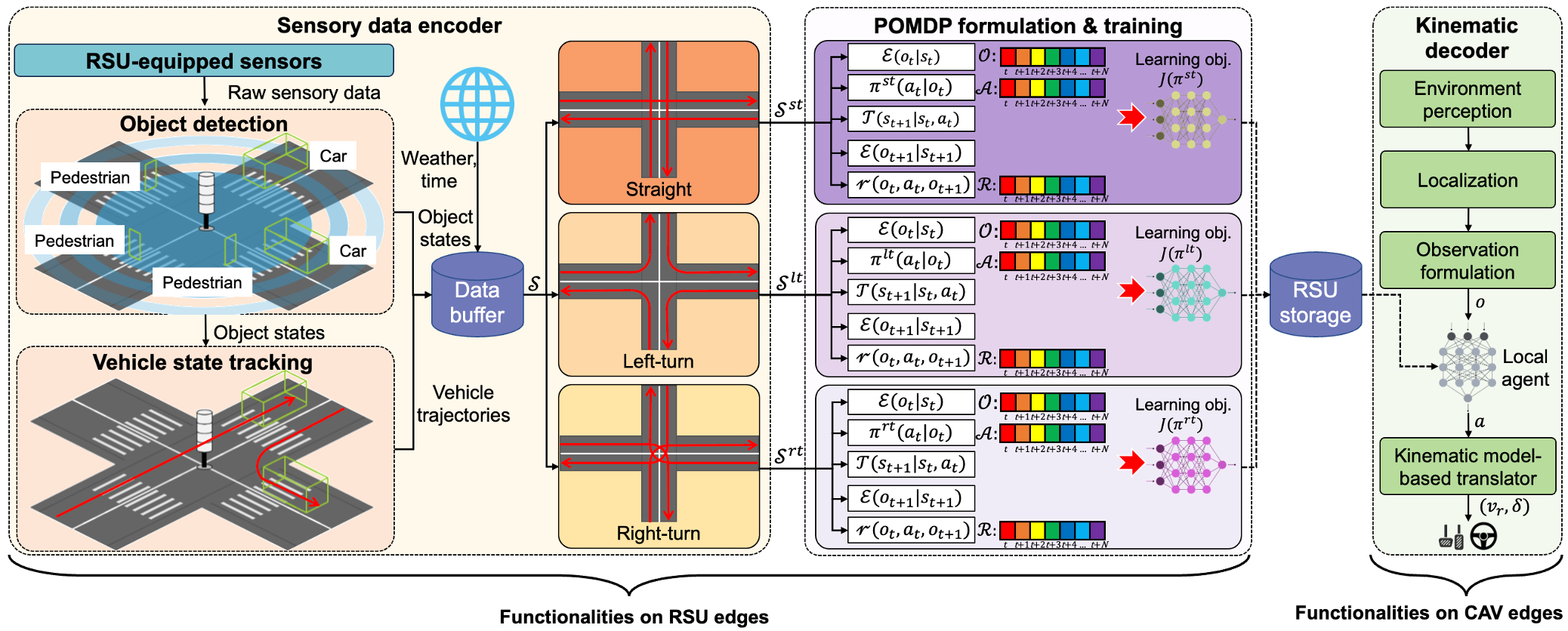}}
    \caption{An overview of offline RL framework across RSU edge, CAV edge, and cloud plane.}
    \label{fig: fra}
\end{figure*}

As shown in Fig.~\ref{fig: fra}, the design of the offline RL framework also includes three key components: RSU edge, CAV edge, and the central cloud, aiming to realize the concept of LDT-assisted hybrid autonomous driving system depicted in Fig.~\ref{fig: sys}. This framework is developed with the following considerations:

\begin{itemize}
    \item We use RSU-installed sensors and edge computing devices to formulate an object-level environmental perception, where the different non-stationary traffic participants, e.g., vehicles, pedestrians, and cyclists, can be identified and their immediate states are obtained. Thus, an LDT can be realized on the RSU edge.
    \item RSU edge is designed to extract passing-by vehicles' behaviors and learn how human drivers maneuver in the local area. Then generate various \textit{local agents} under different conditions, including driving directions, time periods, and weather types.
    \item The maneuvering decoder, i.e., model inference, is placed on the vehicle edge to generate control signals, considering the safety-critical nature of vehicle maneuvering.
\end{itemize}

Therefore, our proposed offline RL framework effectively fulfills the conceptual design, unleashing the resources and capabilities of cloud and edge computing.

\subsection{Sensory Data Encoder}

The dynamic intersection environment is continuously monitored by RSU, which is capable of recognizing, detecting, and tracking all traffic participants within sensory range, covering the entire intersection area. We define $\mathcal{P}_t = \{p_{1}, p_{2}, \ldots, p_{N}\}$ as the set representing all detected traffic participants within the intersection at time step $t$. For each participant $p_k$ in $\mathcal{P}_t$, its state is defined as follows:

\[\boldsymbol{s}_{t,k} = [c_k, l_k, w_k, x_k, y_k, \theta_k, v_k, v_{\textrm{lon},k}, v_{\textrm{lat},k}],\]

\noindent where $c_k$ represents the category of the detected object, which could be a pedestrian, a vehicle, or a cyclist, while $l_k$ and $w_k$ represent its length and width. $x_k$ and $y_k$ capture the absolute position coordinates of the center point of $p_k$. The heading angle, i.e., yaw angle, is represented by $\theta_k$. Variables $v_k$, $v_{\textrm{lon},k}$, and $v_{\textrm{lat},k}$ represent velocity, longitudinal velocity, and lateral velocity of $p_k$, respectively. The set $\mathcal{S}_t = \{\boldsymbol{s}_{t,1}, \boldsymbol{s}_{t,2}, \ldots, \boldsymbol{s}_{t,N}\}$ represents the state space at time step $t$ describing the states of all traffic participants detected by RSU.

Based on real-time object detection, we track the vehicle trajectories to identify their direction of movement and categorize vehicles into three types: straight, left-turn, and right-turn. Thus, the state $\mathcal{S}$ can be divided into three subsets, i.e., $\mathcal{S}^{st}$, $\mathcal{S}^{lt}$, and $\mathcal{S}^{rt}$. This behavior classification ensures the directional diversity of \textit{local agents} so that no matter which road an autonomous vehicle moves in and plans to move out, it can be provided with a feasible agent to guide it in an expected moving direction.

\subsection{Formulation of POMDP}

Since the amount of detected traffic participants $N$ fluctuates over time, it is impractical to include all objects within the observation, particularly for those areas with heavy traffic. Thus, we use partially observed information to determine vehicles' actions. Such a problem can be formulated as a POMDP \cite{levine2020offline}. Offline RL in the context of a POMDP is formally described by a tuple $M = (\mathcal{S}, \mathcal{A}, \mathcal{O}, \mathcal{T}, \mathcal{E}, \mathcal{R}, \gamma)$ where $\mathcal{S}$, $\mathcal{A}$, and $\mathcal{O}$ represent the spaces of states, available actions, and observations. $\mathcal{T}(\boldsymbol{s}_{t+1}| \boldsymbol{s}_t, \boldsymbol{a}_t)$ defines the probability of transitioning to state $\boldsymbol{s}_{t+1}$ from state $\boldsymbol{s}$ after taking action $\boldsymbol{a}_t$ that describes the dynamics of the system. $\mathcal{E}(\boldsymbol{o}_t| \boldsymbol{s}_t)$ is an emission function defining the distribution from $\boldsymbol{s}_t$ to $\boldsymbol{o}_t$. $\mathcal{R}$ is the reward expressed with $r(\boldsymbol{o}_t, \boldsymbol{a}_t, \boldsymbol{o}_{t+1})$ and $\gamma \in (0 ,1]$ is a scalar discount factor.

\begin{figure}[t]
    \centerline{\includegraphics[width=0.40\textwidth]{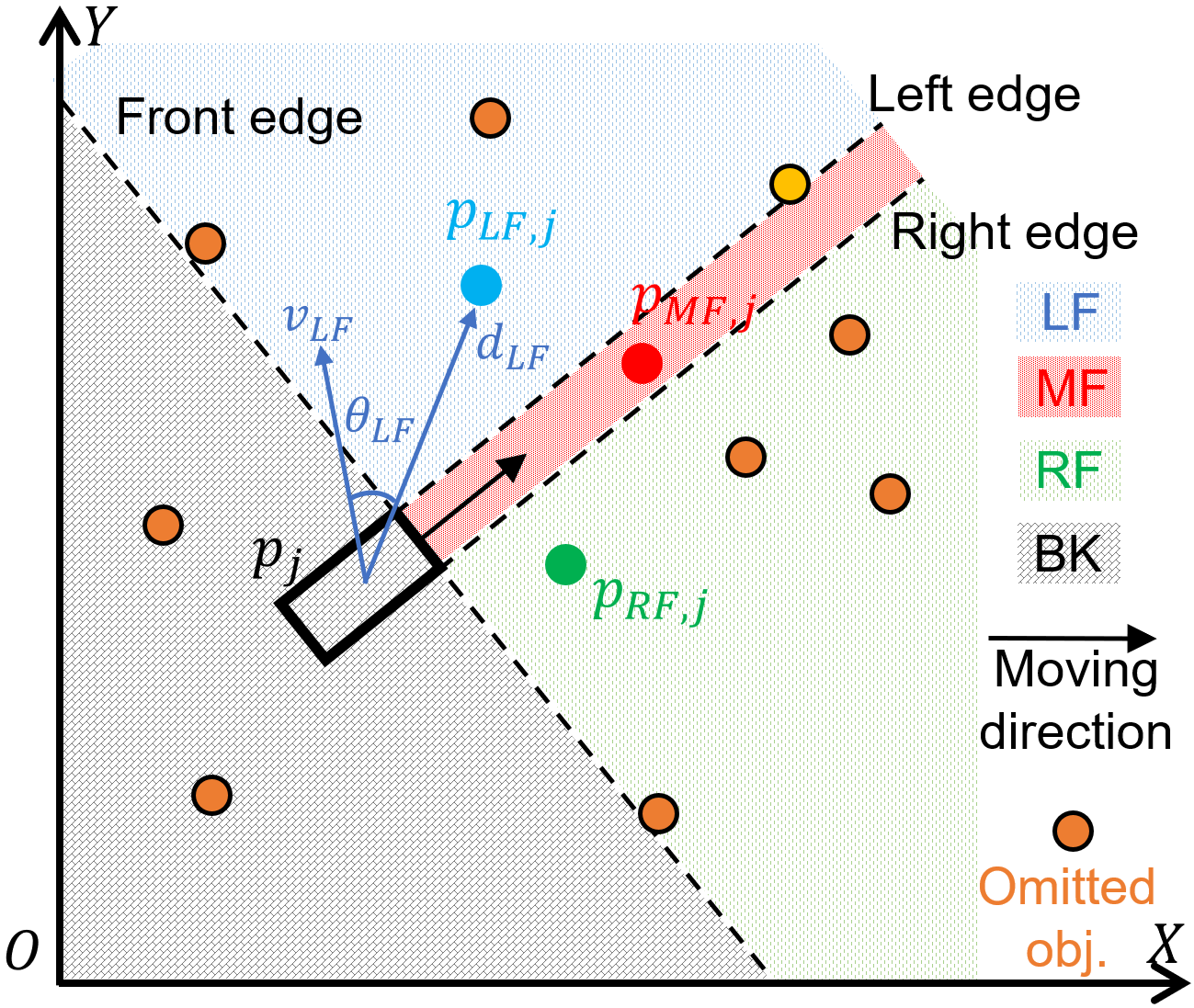}}
    \caption{Illustration of observation space, as well as relative distances and velocity between observed vehicle with its surroundings.}
    \label{fig: obs}
\end{figure}

\textbf{Observation:} At a specific time step $t$, vehicle $p_{j} \in \mathcal{P}_t$ will be passing through the intersection, here, $c_j$ is \textrm{`Vehicle'}. The obstacles in the surrounding area, particularly along the vehicle's heading direction, are critical for road safety. We divide the surrounding area of the vehicle $p_{j}$ into ``left-front'' (LF), ``middle-front'' (MF), ``right-front'' (RF), and ``back''(BK) areas, based on the vehicle’s frontal and lateral boundaries as depicted in Fig.~\ref{fig: obs}.

In the LF, MF, and RF areas, the closest objects to the vehicle $p_{j}$ are denoted by $p_{\textrm{LF}, j}$, $p_{\textrm{MF}, j}$, and $p_{\textrm{RF}, j}$, respectively, as shown in Fig.~\ref{fig: obs}. At any given time $t$, the observation $\boldsymbol{o}_t \in O$ is given by:


\[\boldsymbol{o}_t = [\boldsymbol{s}_{j}, \boldsymbol{s}_{\textrm{LF}, j}, \boldsymbol{s}_{\textrm{MF}, j}, \boldsymbol{s}_{\textrm{RF}, j}],\]

\noindent where $\boldsymbol{s}_{j}$ is the state of vehicle $p_{j}$, and $\boldsymbol{s}_{\textrm{LF}, j}$, $\boldsymbol{s}_{\textrm{MF}, j}$, and $\boldsymbol{s}_{\textrm{RF}, j}$ represent the states of the nearest objects within the LF, MF, and RF areas, respectively. The design of the observation $\boldsymbol{o}_t$ is based on the following considerations: \textit{(i)} Human drivers make decisions based on their visual field, which excludes objects far away from ego vehicle or hidden in blind spots. Thus, from a causality perspective, we only consider visible and closest objects since they directly influence the decision-making of human drivers; \textit{(ii)} Our system fuses onboard sensing and RSU perception data, ensuring that the CAV can detect and react to objects as soon as they enter the observation space, and \textit{(iii)} given the variable amount of detected objects, considering only the nearest objects ensures a practical and computationally efficient observation space while still capturing the most immediate and impactful elements for safe driving.

\textbf{Action:} To effectively accommodate the diverse mechanical constraints of different vehicles, such as baseline length, steering limits, and acceleration capabilities, action $\boldsymbol{a}$ should be designed considering three requirements: \textit{(i)} capabilities to reflect strategic driving decisions in the traffic environment, \textit{(ii)} it can be derived from the vehicle's state $s_{j}$, and \textit{(iii)} can be translated into the control space for different autonomous vehicles. Specifically, for the vehicle $p_{j}$ at time $t$, the action $\boldsymbol{a}_t$ will be:

\begin{equation}
    \boldsymbol{a}_t = [v_{x, j}, v_{y, j}, \dot{\psi}_{j}],
    \label{equ: a_t}
\end{equation}

\noindent where $v_{x}$ and $v_{y}$ represent the vehicle's velocity components along the $X$-axis and $Y$-axis, respectively, and $\dot{\psi}$ represents the yaw rate. Here, the origin of the coordinate system is the location of the RSU. As such, the actions of all vehicles are represented in the same coordinate system. Thus the projection of velocity $v_{x}$ and $v_{y}$ onto this coordinate system becomes an insightful indicator of the vehicle's driving strategy.

\textbf{Reward:} To effectively evaluate the performance of driving strategies within our framework, the reward function is constructed as a linear combination of three critical components: road safety, traveling efficiency, and deviation from the road centerline. 

\begin{equation}
r = \alpha_1 r_{\textrm{safety}} + \alpha_2 r_{\textrm{effi}} + \alpha_3 r_{\textrm{dev}},
\label{equ: reward}
\end{equation}

\noindent where $\alpha_1$,  $\alpha_2$, and  $\alpha_3$ serve as weighting coefficients.

The safety component $r_{\textrm{safety}}$ is defined as the potential time-to-collision (TTC) between the vehicle and the three closest objects. For example, considering the closest object on the LF side of the vehicle $p_{\textrm{LF}, j}$, we use $d_{\textrm{LF}, j}$ and $v_{\textrm{LF}, j}$ to represent the relative distance and relative velocity between the $p_{j}$ and $p_{\textrm{LF}, j}$, and $\theta_{\textrm{LF}, j}$ to define the angle between $d_{\textrm{LF}, j}$ and $v_{\textrm{LF}, j}$, as shown in Fig.~\ref{fig: obs}. In a short time interval, it can be assumed that the vehicle and the object do not change their velocity, so the TTC can be expressed as:

\begin{equation}
    \tau_{\textrm{LF},j} = 
    \begin{cases}
        \frac{d_{\textrm{LF}, j}}{v_{\textrm{LF}, j} \cos{\theta_{\textrm{LF}, j}}}, & \text{if } \theta_{\textrm{LF} ,j} \in [0, \frac{\pi}{2}),  \\
        +\infty, & \text{if } \theta_{\textrm{LF}, j} \in [\frac{\pi}{2}, \pi].
    \end{cases}
\label{equ: ttc}
\end{equation}

When the value of $\theta_{\textrm{LF}, j}$ is smaller than $\pi/2$, we calculate the potential TTC based on relative distance velocity. If $\theta_{\textrm{LF}, j}$ is larger than $\pi/2$, indicating that $p_{j}$ and $p_{\textrm{LF}, j}$ are moving away from each other, then the TTC is set to infinity.
Similar to the calculation of $\tau_{\textrm{LF}, j}$ in (\ref{equ: ttc}), we can obtain $\tau_{\textrm{MF}, j}$ and $\tau_{\textrm{RF}, j}$ for the closest objects on the MF and RF sides. $r_{\textrm{safety}}$ is then determined by:

\begin{equation}
r_{\textrm{safety}} = \sum_{i = \textrm{LF}, \textrm{MF}, \textrm{RF}} \beta_i \frac{\min [\tau_{i ,j}, \tau_{\textrm{thre}}]}{\tau_{\textrm{thre}}},
\label{equ: safetyreward}
\end{equation}

\noindent where $\tau_{\textrm{thre}}$ is a predefined threshold that serves as a boundary to evaluate the urgency of potential collisions. The value of $\tau_{\textrm{thre}}$ is set to 3 seconds based on the Japanese traffic rule \cite{sofa_drivers_guide}. $\beta_i$ is a parameter that balances the importance of the three nearest objects. For example, for a vehicle making a left turn, its expected trajectory is in the LF region, thus $\beta_{\textrm{LF}}$ is assigned a larger value, while $\beta_{\textrm{RF}}$ is with a smaller value, i.e., $\beta_{\textrm{LF}} > \beta_{\textrm{MF}} > \beta_{\textrm{RF}}$. As for a straight-moving vehicle, $\beta_{\textrm{MF}}$ is assigned a larger value, i.e., $\beta_{\textrm{MF}} > \beta_{\textrm{LF}} = \beta_{\textrm{RF}}$.

The efficiency reward component $r_{\textrm{effi}}$ is defined as the optimal travel speeds normalized by the maximum speed,

\[
r_{\textrm{effi}} = 
    \begin{cases}
        \frac{v_{j}}{v_{\textrm{max}}}, & \text{if } v_{j} \leq v_{\textrm{max}},  \\
        - \frac{v_{j} - v_{\textrm{max}}}{v_{\textrm{max}}}, & \text{if } v_{j} > v_{\textrm{max}}, 
    \end{cases}
\]

\noindent where $v_{\textrm{max}}$ represents the speed limit. When $0 \leq v_{j} \leq v_{\textrm{max}}$, $r_{\textrm{effi}}$ is positive to encourage velocity maintenance near the speed limit. When $v_{j} > v_{\textrm{max}}$, $r_{\textrm{effi}}$ becomes negative for the penalty of overspeeding.

Lastly, the reward for minimizing deviation from the lane's centerline $r_{\textrm{dev}}$, is designed to incentivize precise lane-keeping by penalizing deviations from the centerline:

\[
r_{\textrm{dev}} = 1-\frac{2\Delta}{w},
\]

\noindent where $\Delta$ and $w$ represent the lateral displacement from the centerline and the average lane width, where $w$ can be obtained based on static high-definition (HD) maps.

As for the training processes, we introduce the twin delayed deep deterministic policy gradient with behavior cloning (TD3+BC) algorithm \cite{fujimoto2021minimalist} to obtain offline policy $\pi$, since \textit{(i)} it effectively balances exploration and exploitation by combining RL with behavior cloning (BC), ensuring robust policy learning from pre-collected driving data, \textit{(ii)} it reduces overestimation bias in Q-value estimates, which is crucial for accurately evaluating actions in the offline setting of local driving data, and \textit{(iii)} it is computationally efficient and easy to implement, which is essential for training \textit{local driving agents} with limited computational resource of RSU edge. Basically, TD3+BC’s policy $\pi$ is updated with the deterministic policy gradient with a BC term:

\begin{equation}
\pi = \arg\max_{\pi} \mathbb{E}_{(\boldsymbol{s},\boldsymbol{a}) \sim D} \left[ \lambda Q(\boldsymbol{s}, \pi(\boldsymbol{s})) - (\pi(\boldsymbol{s}) - \boldsymbol{a})^2 \right],
\label{equ: policy}
\end{equation}

\noindent where $Q(\boldsymbol{s}, \pi(\boldsymbol{s}))$ represents the Q-value function, and $\lambda$ is a hyperparameter that controls the weight of the Q-value estimation against the behavior cloning (BC) term.

\subsection{Kinematic Decoder for Vehicle Maneuvering}

\begin{figure}[t]
    \centerline{\includegraphics[width=0.40\textwidth]{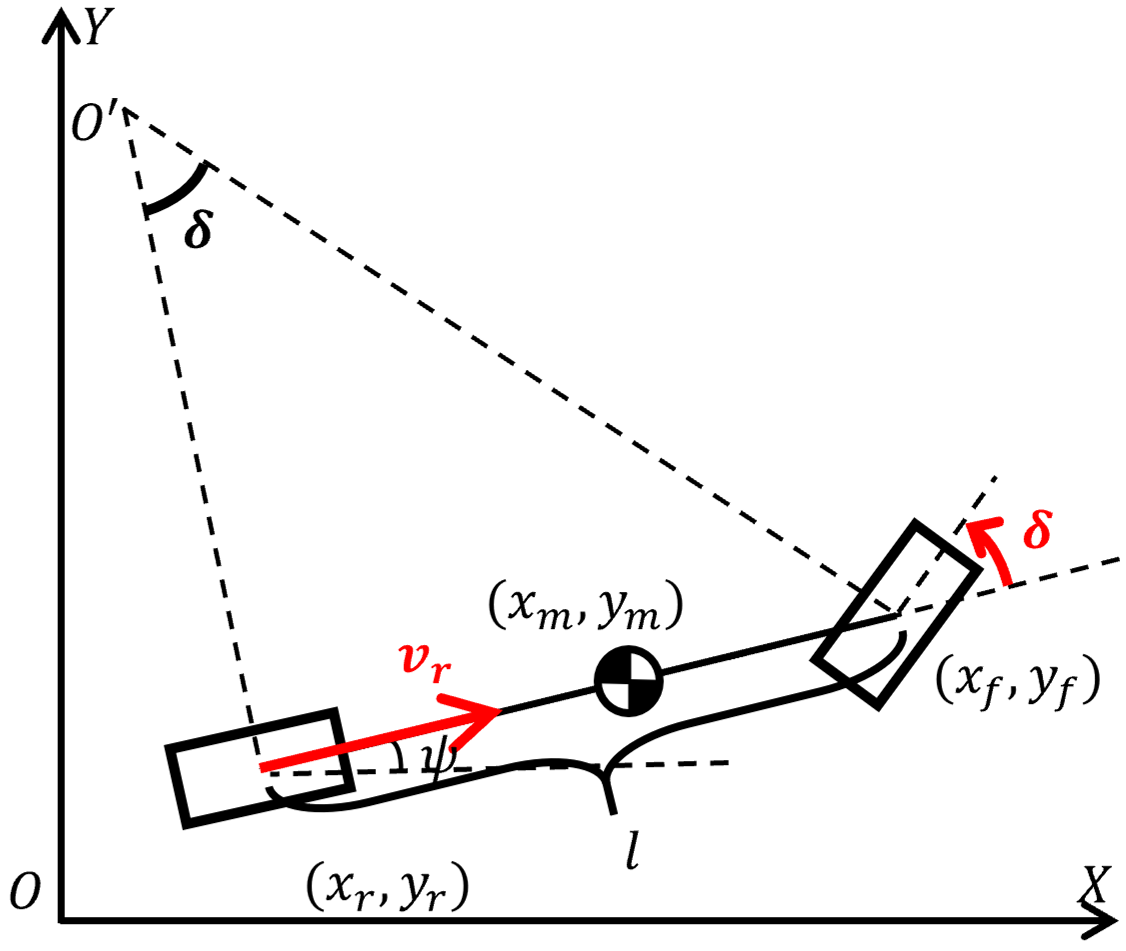}}
    \caption{Illustration of kinematic bicycle model.}
    \label{fig: kine}
\end{figure}

In order to use \textit{local agents} on CAV edge, it is necessary to formulate observation space based on environment perception (generating $\boldsymbol{s}_{LF, j}$, $\boldsymbol{s}_{MF, j}$, and $\boldsymbol{s}_{RF, j}$) and localization (reflecting $\boldsymbol{s}_{j}$) at first. Then we can use trained agents to output actions. However, action space in (\ref{equ: a_t}) cannot be directly used for vehicle control, hence we introduce a kinematic bicycle model to simplify car-like vehicle dynamics, as shown in Fig.~\ref{fig: kine}, and to translate the action space into control space. The ego vehicle is modeled with front-wheel steering and rear-wheel drive as an example of the kinematic decoder design. For vehicles with different configurations, such as front-wheel drive, the corresponding kinematic model (e.g., see \cite{huang2013balanced}) should be applied to achieve vehicle control. We use $(x_{\textrm{f}}, y_{\textrm{f}})$, $(x_{\textrm{r}}, y_{\textrm{r}})$ and $(x_{\textrm{m}}, y_{\textrm{m}})$ to represent the coordinates of centers of the front axle, rear axle, and the center point of the baseline. $v_{\textrm{f}}$ and $v_{\textrm{r}}$ are the velocities of front and rear wheels. The yaw angle and steering angle are represented by $\psi$ and $\delta$, respectively. $l_{\textrm{base}}$ is the length of the vehicle baseline. At the center of the rear axle, the velocity $v_{\textrm{r}}$ can be expressed with its projections on the $X$-axis and $Y$-axis:

\begin{equation}
v_{\textrm{r}} = \dot{x}_{\textrm{r}}\cos{\psi} + \dot{y}_{\textrm{r}}\sin{\psi}.
\label{equ: D1}
\end{equation}

We use the position of the middle point to express the position of the rear axle:

\begin{equation}
\left\{
\begin{aligned}
x_{\textrm{r}} &= x_{\textrm{m}} - \frac{l_{\textrm{base}}}{2}\cos{\psi}, \\
y_{\textrm{r}} &= y_{\textrm{m}} - \frac{l_{\textrm{base}}}{2}\sin{\psi}.
\end{aligned}
\right.
\label{equ: D2}
\end{equation}

The differential form of (\ref{equ: D2}) is:

\begin{equation}
\left\{
\begin{aligned}
\dot{x}_{\textrm{r}} &= \dot{x}_{\textrm{m}} + \frac{l_{\textrm{base}}\dot{\psi}}{2}\sin{\psi}, \\
\dot{y}_{\textrm{r}} &= \dot{y}_{\textrm{m}} - \frac{l_{\textrm{base}}\dot{\psi}}{2}\cos{\psi}.
\end{aligned}
\right.
\label{equ: D3}
\end{equation}

Here, we note that $\dot{x}_{\textrm{m}}$ and $\dot{y}_{\textrm{m}}$ correspond to the $v_x$ and $v_y$ in the action space. Based on (\ref{equ: D1}) and (\ref{equ: D3}), the rear tire velocity can be expressed with the $v_x$ and $v_y$:

\begin{equation}
v_{\textrm{r}} = v_x\cos{\psi} + v_y\sin{\psi}.
\label{equ: D4}
\end{equation}

At the center of the front and rear axles, the kinematic constraints are:

\begin{equation}
\left\{
\begin{aligned}
&\dot{x_{\textrm{f}}}\sin{(\psi + \delta)} - \dot{y_{\textrm{f}}}\cos{(\psi + \delta)} = 0,\\
&\dot{x_{\textrm{r}}}\sin{\psi} - \dot{y_{\textrm{r}}}\cos{\psi} = 0.
\end{aligned}
\right.
\label{equ: D5}
\end{equation}

Based on (\ref{equ: D1}) and~(\ref{equ: D5}), we can obtain:

\begin{equation}
\left\{
\begin{aligned}
\dot{x}_{\textrm{r}} &= v_{\textrm{r}}\cos{\psi}, \\
\dot{y}_{\textrm{r}} &= v_{\textrm{r}}\sin{\psi}.
\end{aligned}
\right.
\label{equ: D6}
\end{equation}

According to the relative positions of the front and rear axles, the position of the front axle is:

\begin{equation}
\left\{
\begin{aligned}
x_{\textrm{f}} &= x_{\textrm{r}} + l_{\textrm{base}}\cos{\psi},\\
y_{\textrm{f}} &= y_{\textrm{r}} + l_{\textrm{base}}\sin{\psi}.
\end{aligned}
\right.
\label{equ: D7}
\end{equation}

Based on (\ref{equ: D5})-(\ref{equ: D7}), the steering angle can be expressed with vehicle yaw rate:

\begin{equation}
\delta = \arctan{\frac{l_{\textrm{base}}\dot{\psi}}{v_{\textrm{r}}}}.
\label{equ: D8}
\end{equation}

For the autonomous driving motion control, $(v_{\textrm{r}}, \delta)$ is regarded as the control space. According to (\ref{equ: D4}) and~(\ref{equ: D8}), we translate the action space to the control space, thereby facilitating motion control for autonomous vehicles.

\section{Real-world Prototype and Experiment Settings}
In this section, we explain how to implement our designed system in a real-world prototype, from the perspective of both hardware and software deployments. Then, we introduce some important settings in HiL and PoC experiments.

\subsection{Hardware Deployment and Software Installation}

\begin{figure*}[!t]
    \centerline{\includegraphics[width=1\textwidth]{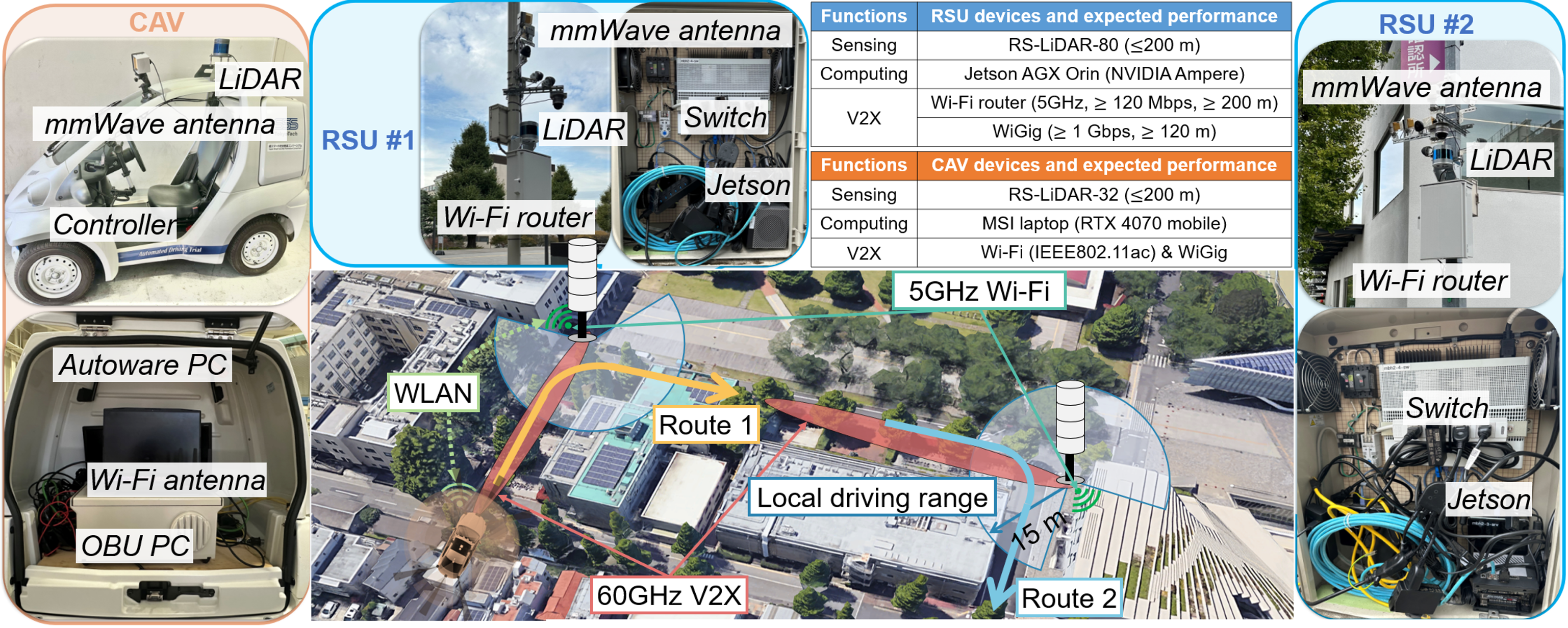}}
    \caption{An overview of the hybrid autonomous driving testing field with specifications of hardware deployment.}
    \label{fig: imp}
\end{figure*}

\begin{figure}[!t]
    \centerline{\includegraphics[width=0.48\textwidth]{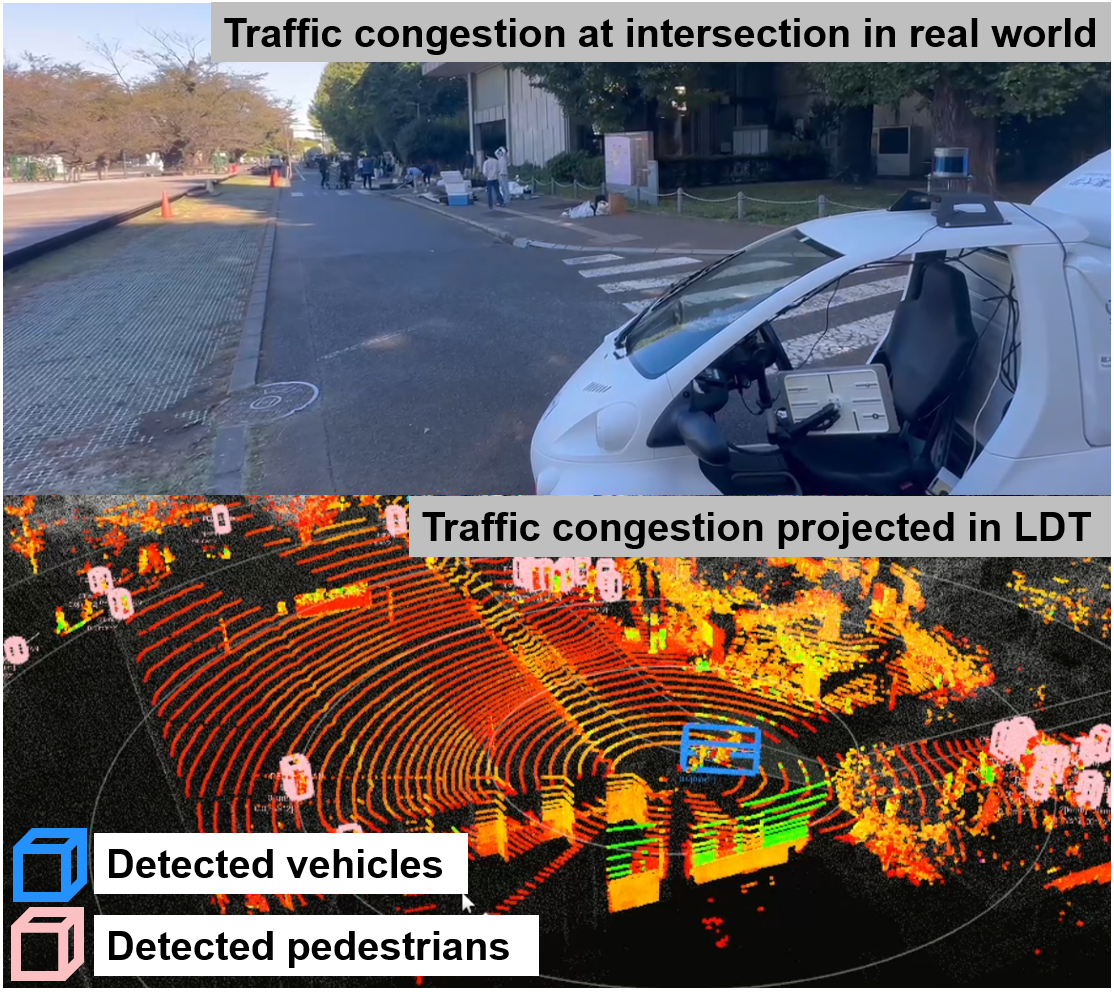}}
    \caption{An illustration of congestion at an intersection in the real world and LDT.}
    \label{fig: cong}
\end{figure}

According to the system design and the requirements in Section II, our hybrid autonomous driving testing field is designed as illustrated in Fig.~\ref{fig: imp}. Within this setting, we install two RSUs at two different intersections. Each RSU is equipped with an 80-layer LiDAR and an NVIDIA Jetson. The LiDAR captures raw data, i.e., dynamic point clouds, from the physical environment. The Jetson is tasked with executing various functional modules, such as object detection and tracking, thereby creating object-level LDT at the RSU edge. As shown in Fig.~\ref{fig: cong}, we demonstrate an example of intersection congestion in the testing field and its corresponding LDT. A CAV is mounted with a 32-layer LiDAR to monitor its surroundings. An Autoware PC processes the LiDAR data, enabling environmental perception, localization, motion planning, and motion control. The control signals are then transmitted to the onboard unit (OBU) to facilitate autonomous driving. The communications system in the hybrid autonomous driving platform is built on a heterogeneous V2X network. For perception information sharing, we employ Wi-Fi modules working at 5~GHz to let CAV receive CP data from RSUs. As for \emph{local driving agents} transfer, given the substantial data size of AI models, we use 60~GHz mmWave communications to enable efficient and high-speed data transmission.

Our platform's software system is primarily based on Autoware.Universe \cite{autoware2022autoware} and version 2 of Robot Operating System (ROS2). Autoware supports the detection and tracking of traffic participants within the RSU sensor range. The detection module employs the CenterPoint framework \cite{yin2021center}, which can detect, identify, and visualize 3D objects from LiDAR point clouds in real time. Subsequently, a multi-object tracker module assigns IDs to the detected objects so that we can record each object's historical trajectory and classify them based on moving directions. As for the autonomous driving system, we also apply a Normal Distributions Transform (NDT) algorithm, OpenPlanner, and a pure pursuit algorithm for localization, motion planning, and motion control modules, respectively. Detailed information about the hardware and software systems in the test field can be found in \cite{wang2024smart}.

\subsection{Simulation and Field Trial Setups}

Following the discussion in Section III.B and III.C, we first utilize two RSUs to monitor their surrounding traffic flows and generate right-turn \textit{local driving agents}, named RSU \#1 local agent and RSU \#2 local agent. Then, we conduct PoC tests and HiL simulations to validate the proposed system. As depicted in Fig.~\ref{fig: imp}, we define two right-turn routes (Route \#1 and Route \#2) that respectively pass by RSU \#1 and RSU \#2-installed intersections. We also define a `local driving range' (15 m). When CAV enters the `local driving range', CAV will adopt \textit{local driving agents} as a replacement for its inherent motion planning and motion control modules. The local driving range of 15 m is determined based on the road topology of the test field and vehicle speed limitation of 20 km/h, where the 15 m range is sufficient to cover the entire intersection. For larger intersections or scenarios involving higher vehicle speeds, the local driving range should be redefined accordingly. In addition, to test the autonomous driving performance in different traffic conditions, we categorize the traffic into 3 types: low density (the number of objects within the intersection area is less than 3), middle density (ranging from 3 to 6), and high density (larger than 6). Some detailed settings in the simulation and field trial are described as follows and summarized in Table III.

\subsubsection{HiL simulation} In order to repeatedly evaluate the performance of our system in autonomous driving, we conducted HiL simulations using real traffic data perceived by RSUs and a simulated vehicle. We utilized 3 types of agents (RSU \#1 local agent, RSU \#2 local agent, and the inherent Autoware agent) to drive the simulated CAV through Route \#1 and Route \#2 under 3 different traffic conditions (low, middle, and high density). Each scenario was tested 10 times, resulting in a total of 180 experiments.

\subsubsection{PoC field trial} Our experimental environment involves public traffic and the intersection at Route \#2 presents highly complex traffic conditions. Therefore, for safety reasons, we conduct field tests only on Route \#1. Here, we define a `communications distance' (30 m), within which the V2I communications is activated that RSU will disseminate \emph{local driving agents} through mmWave communications and provide processed data CP services through the WLAN network. We conducted 9 experiments in total where we employed 3 types of agents (RSU \#1 local agent, RSU \#2 local agent, and the inherent Autoware agent) to navigate the CAV through Route \#1 under 3 different traffic conditions (low, middle, and high density).

\begin{table}[t]
\centering
\renewcommand{\arraystretch}{1.2}  

\caption{PoC and HiL setups.}
\begin{tabular}{cc}
\hline
\hline
\rowcolor[HTML]{FFFC9E} 
\multicolumn{1}{c|}{\cellcolor[HTML]{FFFC9E}\textbf{Specifications}}                                                        & \textbf{Value/Description}                                                                                                                           \\ \hline
\rowcolor[HTML]{EFEFEF} 
\multicolumn{1}{c|}{\cellcolor[HTML]{EFEFEF}Local driving range}                                                   & 15 m (CAV adopts local agents)                                                                                                                                       \\
\multicolumn{1}{c|}{\begin{tabular}[c]{@{}c@{}}Autonomous \\ vehicle actions\end{tabular}}                         & \begin{tabular}[c]{@{}c@{}}CAV adopts local driving agents when \\ entering the defined local driving range\end{tabular}                    \\
\rowcolor[HTML]{EFEFEF} 
\multicolumn{1}{c|}{\cellcolor[HTML]{EFEFEF}\begin{tabular}[c]{@{}c@{}}Traffic density \\ conditions\end{tabular}} & \begin{tabular}[c]{@{}c@{}}Low (\textless{}3 objects), Middle (3-6 objects), \\ and High (\textgreater{}6 objects)\end{tabular}             \\
\multicolumn{1}{c|}{Agent types}                                                                                   & \begin{tabular}[c]{@{}c@{}}RSU \#1 Local Agent, RSU \#2 Local Agent, \\ and Inherent Autoware Agent\end{tabular}                            \\ \hline
\rowcolor[HTML]{FFFC9E} 
\multicolumn{2}{c}{\cellcolor[HTML]{FFFC9E}\textbf{PoC experiments}}                                                                                                                                                                                                      \\ \hline
\rowcolor[HTML]{EFEFEF} 
\multicolumn{1}{c|}{\cellcolor[HTML]{EFEFEF}\begin{tabular}[c]{@{}c@{}}Field trial \\ count\end{tabular}}          & \begin{tabular}[c]{@{}c@{}}9 tests (3 tests per scenario, conducted on \\ Route \#1 only, using 3 types of agents)\end{tabular}             \\
\multicolumn{1}{c|}{Used RSU(s)}                                                                                   & RSU \#1                                                                                                                                     \\
\rowcolor[HTML]{EFEFEF} 
\multicolumn{1}{c|}{\cellcolor[HTML]{EFEFEF}Driving route(s)}                                                      & Route \#1                                                                                                                                   \\
\multicolumn{1}{c|}{\begin{tabular}[c]{@{}c@{}}Communications \\ distance\end{tabular}}                             & 30 m (V2I communications range)                                                                                                              \\ \hline
\rowcolor[HTML]{FFFC9E} 
\multicolumn{2}{c}{\cellcolor[HTML]{FFFC9E}\textbf{HiL simulations}}                                                                                                                                                                                                      \\ \hline
\rowcolor[HTML]{EFEFEF} 
\multicolumn{1}{c|}{\cellcolor[HTML]{EFEFEF}\begin{tabular}[c]{@{}c@{}}Simulation \\ count\end{tabular}}           & \begin{tabular}[c]{@{}c@{}}180 tests (10 tests per scenario, 3 traffic \\ conditions for both routes, using 3 types of agents)\end{tabular} \\
\multicolumn{1}{c|}{Used RSU(s)}                                                                                   & RSU \#1 and RSU \#2                                                                                                                         \\
\rowcolor[HTML]{EFEFEF} 
\multicolumn{1}{c|}{\cellcolor[HTML]{EFEFEF}Driving route(s)}                                                      & Route \#1 and Route \#2                                                                                                                     \\ \hline
\hline
\end{tabular}
\label{tab:setup}
\end{table}

\section{Experiment Results and System Evaluation}
In this section, we first discussed the offline RL training results. Then, to evaluate the crucial KPIs of the implemented hybrid autonomous driving system, we empirically measured several communications metrics in the prototype. Finally, we showed both the HiL and PoC test results in terms of safety, efficiency, and real-time performance to validate the improvement brought to autonomous driving.

\subsection{Performance Comparison of Offline RL Algorithms}

\begin{figure}[t]
\centering
\subfigure[]{
\begin{minipage}[b]{0.45\textwidth}
\includegraphics[width=1\textwidth]{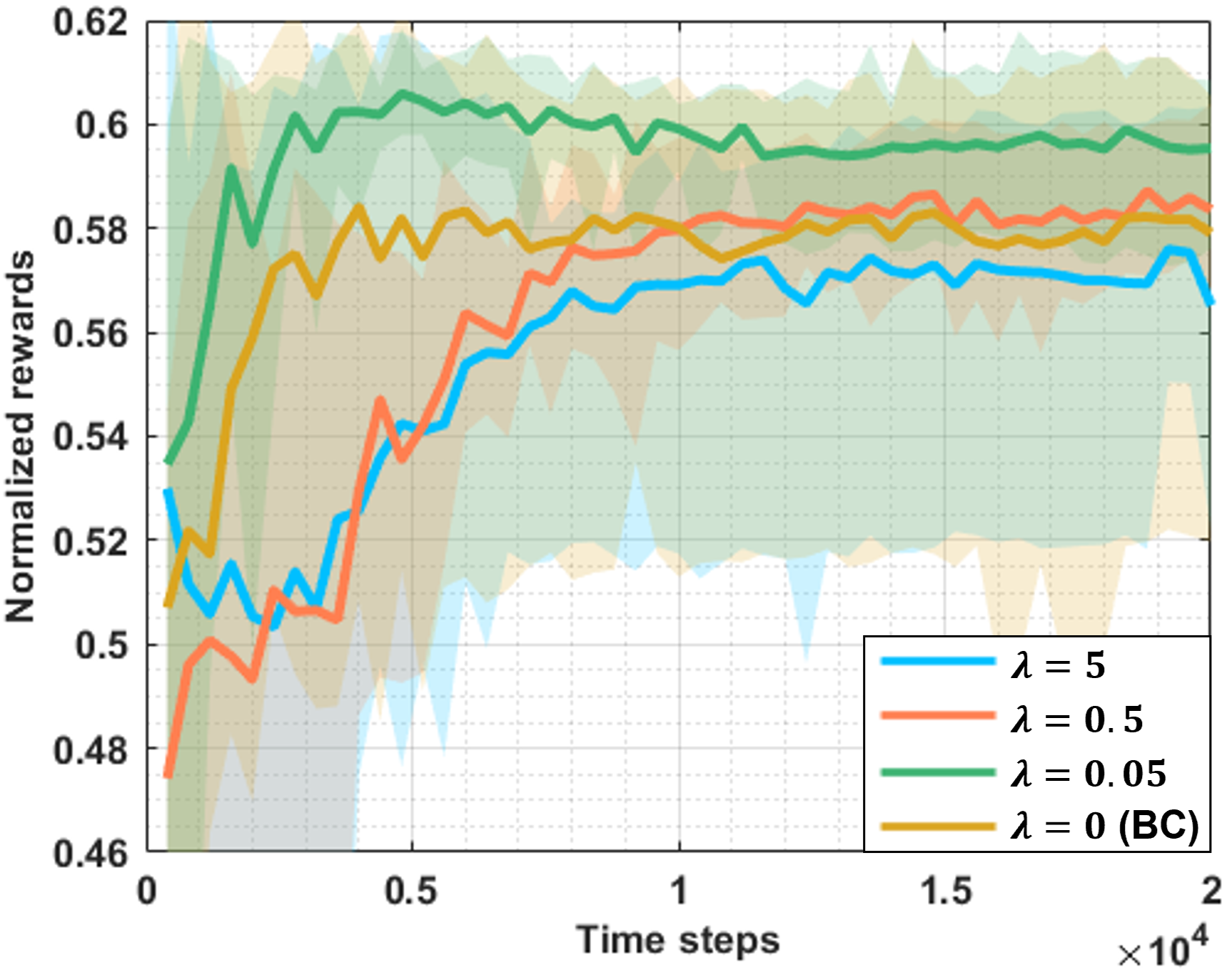} 
\end{minipage}
}
\subfigure[]{
\begin{minipage}[b]{0.45\textwidth}
\includegraphics[width=1\textwidth]{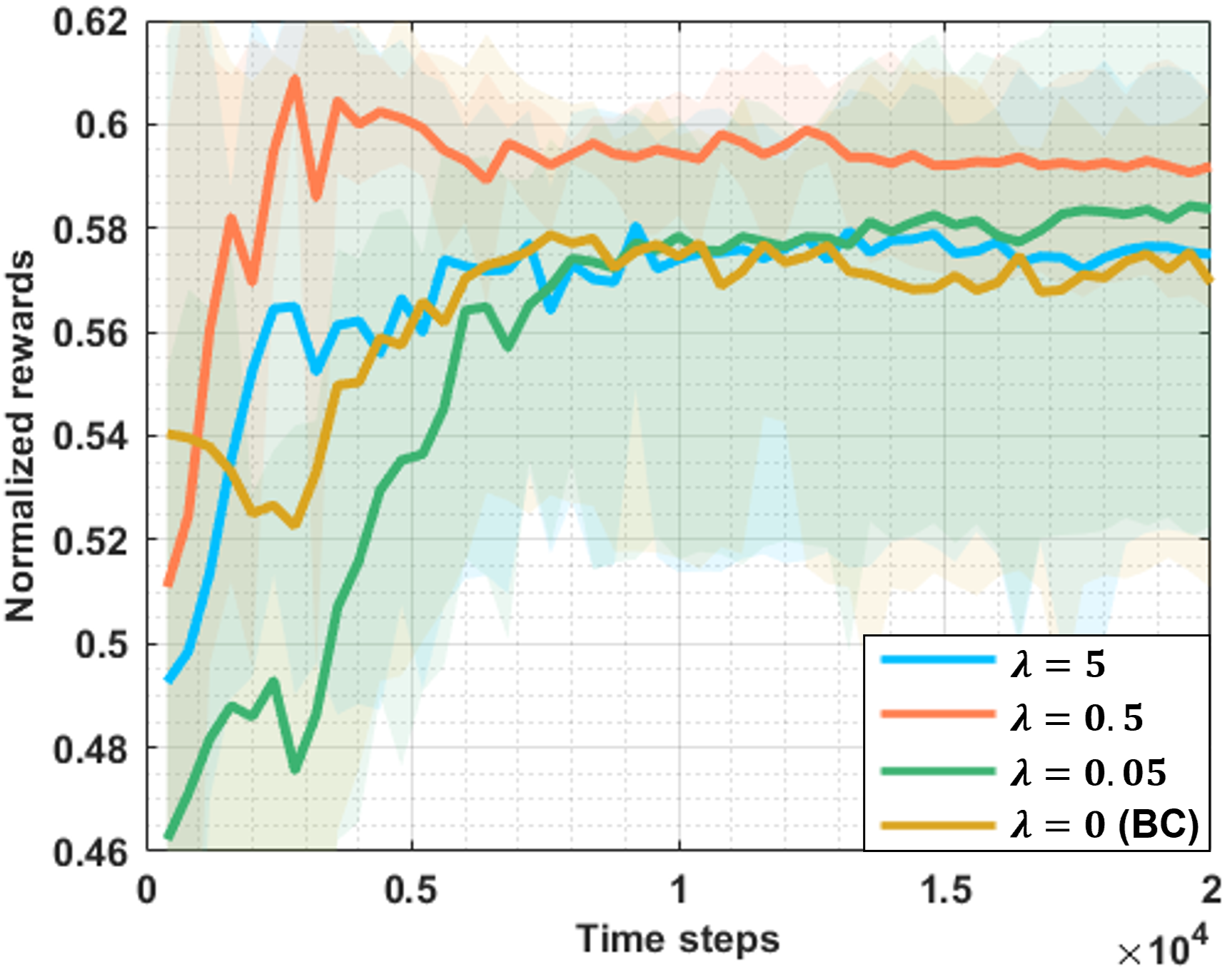} 
\end{minipage}
}
\caption{Offline RL training results on offline traffic data: Normalized rewards by varying $\lambda$ value over 10 random seeds, obtained from (a) RSU \#1-installed intersection traffic, and (b) RSU \#2-installed intersection traffic.}
\label{fig: trainingresult}
\end{figure}

We provide the training performance in terms of normalized rewards in Fig.~\ref{fig: trainingresult}, where the solid lines and dimmed areas represent the mean and confidence interval over 10 random seeds, respectively. The maximum time step to run our customized environment is set to 20,000, with evaluations conducted every 400 time steps. Remarkably, the TD3+BC algorithm can always achieve convergence with varying hyperparameter $\lambda$ in (\ref{equ: policy}), which demonstrates the effectiveness of our dataset and the feasibility of our POMDP formulation. 

For RSU \#1-installed intersection, the convergence result with $\lambda=0.05$ is around 0.59, surpassing that with other $\lambda$ values. As for RSU \#2, the convergence performance is optimal when $\lambda=0.5$. Therefore, in subsequent simulation and field trial tests, models of TD3+BC algorithms trained with $\lambda=0.05$ and $\lambda=0.5$ are regarded as RSU \#1 and RSU \#2 local agents, respectively.

\begin{table}[]
\centering
\renewcommand{\arraystretch}{1.2}  

\caption{Communications performance of agent downloading and CP services.}
\begin{tabular}{c|c|c}
\hline
\hline
\textbf{Services}                                          & \textbf{Agent downloading (mmWave)}                                                    & \textbf{CP (Wi-Fi)}                                      \\ \hline
\rowcolor[HTML]{EFEFEF} 
\begin{tabular}[c]{@{}c@{}}Reliability\\ (\%)\end{tabular}         & - & \begin{tabular}[c]{@{}c@{}}$\ge$ 96.1\\ (97.3 on average)\end{tabular}                                                                     \\ \hline
\rowcolor[HTML]{FFFC9E} 
\begin{tabular}[c]{@{}c@{}}Latency\\ (ms)\end{tabular}     & \begin{tabular}[c]{@{}c@{}}$\leq$ 146\\ (83.9 on average)\end{tabular}  & \begin{tabular}[c]{@{}c@{}}$\leq$ 8.51\\ (6.57 on average)\end{tabular} \\ \hline
\rowcolor[HTML]{EFEFEF} 
\begin{tabular}[c]{@{}c@{}}Data rate\\ (Mbps)\end{tabular} &  \begin{tabular}[c]{@{}c@{}}$\ge$ 711\\ (1240 on average)\end{tabular}  & \begin{tabular}[c]{@{}c@{}}$\ge$ 25.9\\ (32.1 on average)\end{tabular}  \\ \hline
\rowcolor[HTML]{FFFC9E} 
\begin{tabular}[c]{@{}c@{}}Data size\\ (MB)\end{tabular}   & 11$\sim$13                                                                          & -                                                                  \\ \hline
                                                                           \hline
\end{tabular}
\label{tab:commper}
\end{table}

\subsection{System Evaluation Considering Communications Issues}

Some communications metrics are measured in PoC tests, as shown in Table~\ref{tab:commper}. Compared to the required KPIs listed in Table I, Wi-Fi-based CP and mmWave communications-enabled agent downloading services can both meet the specified requirements. As for the agent downloading service, we first measure the data size of agents trained with TD3+BC algorithms, which ranges between 11 and 13 MB, fitting within the defined range of 2 to 40 MB. The data rate achieved during the downloading process is at least 0.711 Gbps, averaging 1.24 Gbps, greatly surpassing the peak data rate requirement of 100 Mbps. The E2E latency for downloading agents is observed to be no more than 146 ms with an average latency of 83.9 ms, which is also well within the acceptable range of 500 to 1000 ms as defined in the 3GPP TR 22.874. Here, we note that mmWave communications may face challenges in extreme environments, such as rain or obstructions, which can affect communications stability. The heterogeneous V2X communications utilized in our system can mitigate such challenges by switching to Wi-Fi-based V2I or cellular networks to ensure reliable communications.

As for CP, since ROS2 facilitates seamless communications among distributed computers within a Local Area Network (LAN), the CAV can easily access all ROS2-defined messages from the RSU edge using user datagram protocol (UDP). In this study, the CAV is designed to subscribe to object detection messages generated in nearby RSUs to facilitate CP services. We evaluate the communications reliability using the observed packet delivery ratio (PDR), which achieves a minimum of 96.1\%, exceeding the 90\% reliability threshold specified for the SSMS use case. Furthermore, the E2E latency and data rate are observed to be less than 8.51 ms and higher than 25.9 Mbps, also well-satisfying the 10 ms latency and 25 Mbps peak data rate requirements. These results confirm that the proposed system supports CP and agent downloading services to fulfill the stringent KPIs for SSMS and AI/ML based automotive networked systems use cases.

\subsection{HiL and PoC Test Results}

\begin{figure*}[t]
\centering
\subfigure[]{
\begin{minipage}[b]{0.31\textwidth}
\includegraphics[width=1\textwidth]{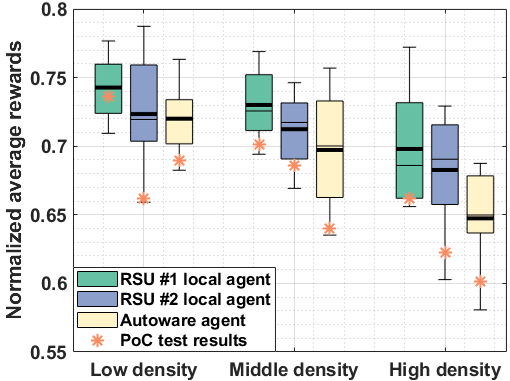} 
\end{minipage}
}
\subfigure[]{
\begin{minipage}[b]{0.31\textwidth}
\includegraphics[width=1\textwidth]{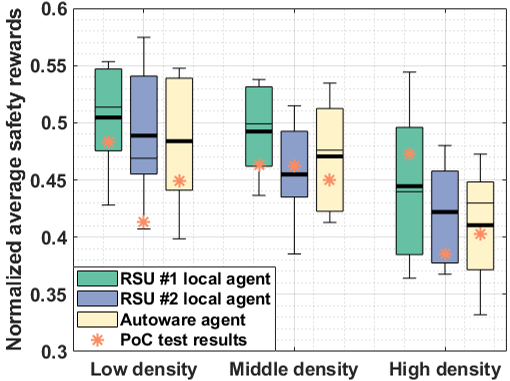} 
\end{minipage}
}
\subfigure[]{
\begin{minipage}[b]{0.31\textwidth}
\includegraphics[width=1\textwidth]{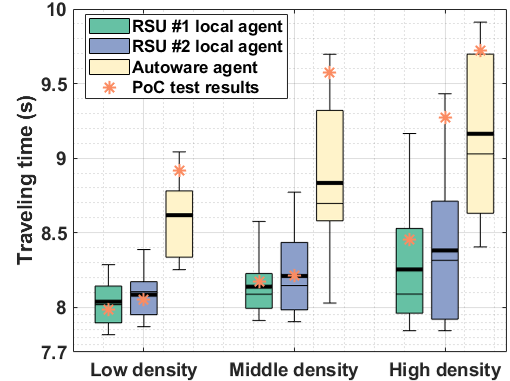} 
\end{minipage}
}
\caption{HiL simulation and PoC test results of applying RSU \#1 local agent, RSU \#2 local agent, and Autoware agent in RSU \#1-installed `local driving range'. (a) Distribution of normalized combined rewards, (b) Distribution of normalized safety rewards, (c) Distribution of traveling time.}
\label{fig: rsu1result}
\end{figure*}

\begin{figure*}[t]
\centering
\subfigure[]{
\begin{minipage}[b]{0.31\textwidth}
\includegraphics[width=1\textwidth]{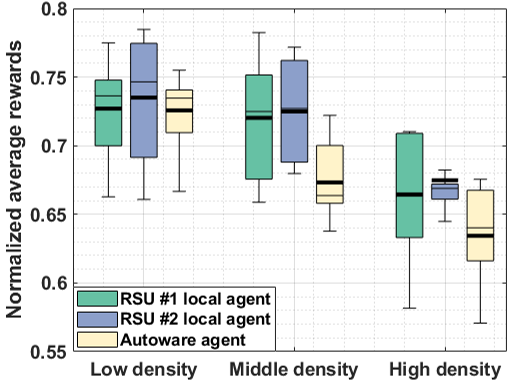} 
\end{minipage}
}
\subfigure[]{
\begin{minipage}[b]{0.31\textwidth}
\includegraphics[width=1\textwidth]{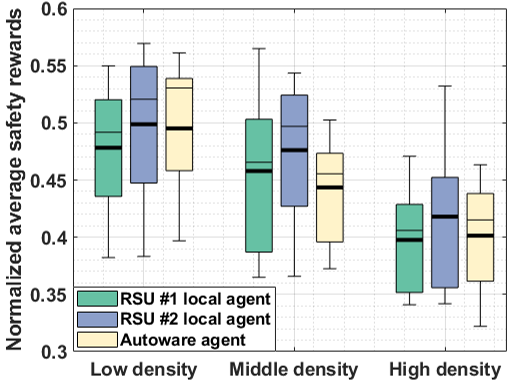} 
\end{minipage}
}
\subfigure[]{
\begin{minipage}[b]{0.31\textwidth}
\includegraphics[width=1\textwidth]{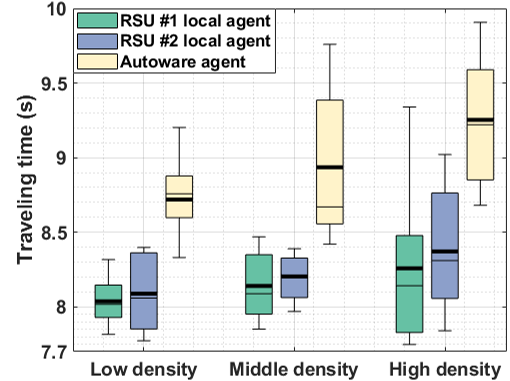} 
\end{minipage}
}
\caption{HiL simulation results of applying RSU \#1 local agent, RSU \#2 local agent, and Autoware agent in RSU \#2-installed `local driving range'. (a) Distribution of normalized combined rewards, (b) Distribution of normalized safety rewards, (c) Distribution of traveling time.}
\label{fig: rsu2result}
\end{figure*}

As discussed in Section IV.B, we conduct both HiL and PoC experiments in RSU \#1-installed intersection, whose results are shown in Fig.~\ref{fig: rsu1result}, where box plots are employed to visually represent the distribution of combined rewards (defined in (\ref{equ: reward})), safety rewards (defined in (\ref{equ: safetyreward})), and traveling time attained by various agents during autonomous driving process, and the mean value is delineated using a bold solid line. We also mark the performance of PoC field tests in Fig.~\ref{fig: rsu1result} as orange star marks. As for RSU \#2-installed intersection, we only showcase the HiL simulation results in Fig.~\ref{fig: rsu2result}. 

According to the data presented in Fig.~\ref{fig: rsu1result}(a), it is evident that RSU \#1 local agent shows better overall performance than RSU \#2 local agent and Autoware agent in Intersection \#1, no matter within simulation or field trial. As illustrated in Fig.~\ref{fig: rsu2result}(a), it also can be found that RSU \#2 local agent always drives the vehicle better than RSU \#1 local agent and Autoware agent in Intersection \#2. Such outcomes validate the effectiveness of our system design and offline RL formulation from the following two perspectives: \textit{(i)} The overall superior performance of the designed local agents, in comparison to the general algorithm, suggests that the proposed system has effectively enhanced autonomous driving at traffic intersections, and \textit{(ii)} The different performance of the two trained agents at two intersections demonstrates the uniqueness of each trained agent, suggesting that models trained on local traffic flow data perform well only when applied to their respective local environments. This indicates that our system effectively captures and learns intersection-specific local strategies from human driving behaviors.

To quantify the evaluation of safety, we employ the safety reward function expressed in (\ref{equ: safetyreward}). Although the Autoware agent adopts a relatively conservative and safety-oriented driving strategy, which refrains from actively overtaking obstacles and instead opts to wait for their leaving, the average safety rewards depicted in Fig.~\ref{fig: rsu1result}(b) and Fig.~\ref{fig: rsu2result}(b) show that the trained agents exhibit safety improvements. For Intersection \#1, compared to the conventional autonomous driving algorithm, i.e., Autoware agent, the local agent exhibits a safety promotion of 5.8\% (on average) and 9.3\% within HiL simulation and PoC experiments, respectively. As for Intersection \#2, its local agent also demonstrates better safety measures with a 4.1\% improvement on average.

For evaluating driving efficiency, we show the distribution of traveling time within defined `local driving range' to offer an intuitive representation of efficiency performance. Fig.~\ref{fig: rsu1result}(c) and Fig.~\ref{fig: rsu2result}(c) illustrate that the local agents exhibit considerable enhancements in driving time reduction. RSU \#1 local agent can reduce travel time by 8.9\% on average according to simulation results, and up to 14.6\% in field tests. In the HiL experiments at Intersection \#2, it can be found that RSU \#2 local agent also decreases the total moving time by 9.1\%.

Finally, we evaluate the real-time performance of proposed local driving agents and some update frequencies of some crucial functional blocks. The CAN controller of our CAV is designed to operate at 10 Hz, requiring the autonomous driving software system to send control signals every 100 ms. The low-level motion control module of inherent Autoware system also operates at 10 Hz, ensuring that the Autoware can control our autonomous vehicle in real-time effectively. For our hybrid autonomous driving system, we replace the motion planning and motion control functions at intersections with an online execution (model inference) of local driving agents. The input information for the model inference includes the vehicle's onboard perception data, processed CP information shared by RSUs, and vehicle localization. All input data updates at a frequency larger than 30 Hz. In addition, the average computation time for model inference is measured at 38.4 ms. Therefore, the output frequency of the local agent can be set to 10 Hz to ensure real-time input and output, thus enabling vehicle control in real time. Consequently, our designed hybrid autonomous driving system meets the real-time control requirements for autonomous driving.

\section{Conclusion}
In this paper, we proposed a novel LDT-assisted hybrid autonomous driving system, specifically designed for traffic intersections to improve road safety and driving efficiency. We addressed the challenge of the intricate and dynamic nature of intersections, where road topology, traffic volume, and complex traffic conditions necessitate a local decision-making approach. By leveraging RSUs equipped with sensory and edge computing capabilities, the system continuously monitors traffic, extracts human driving knowledge, and generates intersection-specific local driving agents through an offline RL framework. This method enables vehicles to adopt local driving agents generated from accumulated driving behavior data, particularly when passing through RSU-installed intersections.

When CAVs pass through RSU-installed intersections, RSUs can provide them not only with real-time CP services to broaden onboard sensory horizons but also with local driving agents to support safe and efficient driving in local areas. Extensive PoC field trials and HiL simulations validate the effectiveness of the system. The results demonstrate that our hybrid system outperforms the conventional Autoware algorithms, achieving average safety improvements of up to about 10\% and travel time reductions of up to about 15\%.

Additionally, the system meets stringent real-time performance requirements, with communications metrics showing peak latencies of 8.51 ms for CP services and 146 ms for local agent downloading, aligning with 3GPP requirements \cite{3gpptr22874, 3gpptr22886}. These results confirm that integrating LDT-based \textit{local driving strategies} can significantly enhance autonomous driving at complex urban intersections. Future work will focus on integrating federated learning and continual learning to further enhance the offline RL model by leveraging distributed data sources.

\bibliographystyle{IEEEtran.bst}
\bibliography{bibliography}

\vfill

\end{document}